\documentclass[sigconf]{acmart}

\usepackage[ruled,vlined,linesnumbered]{algorithm2e}
\usepackage{multirow}
\usepackage{url}
\renewcommand\footnotetextcopyrightpermission[1]{}

\begin{document}

\title{Circle Feature Graphormer: Can Circle Features Stimulate Graph Transformer?}

\author{Jingsong Lv, Hongyang Chen, Yao Qi, Lei Yu}
\affiliation{%
  \institution{Research Center for Graph Computing, Zhejiang Lab}
  \city{Hangzhou}
  \state{Zhejiang Province}
  \country{China}
}
\email{{jingsonglv,hongyang,qiy, yulei}@zhejianglab.com}

\renewcommand{\shortauthors}{Jingsong Lv, et al.}

\begin{abstract}
  In this paper, we introduce two local graph features for missing link prediction tasks on ogbl-citation2. We define the features as Circle Features, which are borrowed from the concept of circle of friends. We propose the detailed computing formulas for the above features. Firstly, we define the first circle feature as modified swing for common graph, which comes from bipartite graph. Secondly, we define the second circle feature as bridge, which indicates the importance of two nodes for different circle of friends. In addition, we firstly propose the above features as bias to enhance graph transformer neural network, such that graph self-attention mechanism can be improved. We implement a Circled Feature aware Graph transformer (CFG) model based on SIEG network, which utilizes a double tower structure to capture both global and local structure features. Experimental results show that CFG achieves the state-of-the-art performance on dataset ogbl-citation2.
\end{abstract}

\ccsdesc[500]{Computing methodologies~Machine learning}

\keywords{Graph Transformer, Graph Neural Networks, Circle Feature, Representation Learning, Link Prediction. }

\maketitle

\section{Introduction}
Link prediction is an important task to find latent relationship between nodes. A traditional method is to utilize the node and edge features or latent node and edge features to predict such missing links. Such node features include degree, centrality level, common neighbors\cite{liben2007link}, curvature\cite{federer1959curvature}, pagerank\cite{page1998pagerank}, simrank~\cite{jeh2002simrank}, and so on. In addition, node related sub network features can also be used to predict missing links. A famous example is triangles between nodes. Furthermore, multi-hop adjacent neighborhood structure information can improve the prediction performance. Inspired by the success of the deep Convolutional Neural Network(CNN)\cite{li2021survey} in the field of computer vision, Graph Neural Network(GNN) \cite{wu2020comprehensive} is widely applied in the task of link prediction. GNN can capture deep receptive field information, which cover more information than local sub-network structure, such that the performance can be greatly improved. Recently, transformer based neural network beats traditional GNN for the capability of capturing global graph information. 

In this paper, we introduce the concept of circle feature, which is borrowed from circle friends. The motivation is that nodes with more common circles have stronger connectiveness. We propose two circle features and give the formal definitions. In addition, We propose a Circle Feature aware Graph transformer(CFG) method for link prediction, which improve the transformer self-attention mechanism. Finally, we conduct experiments on dataset ogbl-citation2, and show that CFG beats the state-of-the-art method for missing link prediction task.

\section{Related Work}
\label{sec:relatedwork}
\subsection{Link Prediction} Approaches for link prediction can be divided into three groups: neighborhood based, factor decomposition based and neural network based. Neighborhood based methods mainly search out neighborhood information in a wide-based or depth-based manner such that nodes similarity can be figured out and used for prediction of missing links, such as common neighbors~\cite{liben2007link},  Adamic-Adar~\cite{adamic2003friends}, SimRank~\cite{jeh2002simrank} and Node2Vec~\cite{grover2016node2vec}. Factor decomposition based methods mainly utilize latent factors to measure nodes similarity, such as PNRL~\cite{wang2017predictive}. In addition,  neural network based methods are used to capture deep hierarchical structure features in an end-to-end manner, such as GCN~\cite{kipf2016semi}, SEAL~\cite{zhang2018link}. 

\subsection{Cycle Enumeration} Enumeration of simple cycles of graphs is a classical computer science problem whose efficient solutions data back to the early 70s, which has many important applications in several domains, ranging from the feedback loops detection in biological networks\cite{klamt2009computing} to the mechanical analysis of chemical structures\cite{sussenguth1965graph}.

The most efficient solution was presented by Johnson\cite{johnson1975finding}, Read and Tarjan\cite{read1975bounds}, and Szwarcfiter and Lauer\cite{szwarcfiter1976search}, which achieve the lowest time complexity bounds for listing simple cycles in directed graphs. But this is not optimal for undirected graphs, an optimal cycle enumeration algorithm for undirected graphs was presented by Birmel\'{e} et al.\cite{birmele2013optimal}. 

Due to the complexity of calculation and the need of practical application, various types of constraints are imposed on the cycles. For instance, long cycles in financial transaction networks are less likely to be associated with money laundering, so hop constraints can be imposed in this situation\cite{qin2019towards}. Cycles of specific length attract much more research, such as, triangles, which has many real-life applications, the most-known among them, is to compute the clustering coefficient\cite{watts1998collective} and the transitivity ratio\cite{luce1949method} of a graph.

\section{Methods}
\label{sec:methods}

\subsection{Problem Definition}
Let $G=(V,E)$ represent a homogeneous, unweighted, undirected graph, where $V$ represents the nodes set of the graph, and $E \subseteq V \times V$ represents the edges set of the graph. Let $N=|V|$ and $M=|E|$. Given an edge $e_{ij} \in E$, it can be represented as node pair $(v_i, v_j)$, where $v_i,v_j \in V$.  Now let $A \in \mathbb{R}^{N \times N}$ represent the corresponding symmetric adjacency matrix, where matrix element $a_{ij}$ indicates whether edge $(v_i, v_j) \in E$. For node features, let $X\in \mathbb{R}^{N \times d}$ represent the feature matrix, where the matrix element $x_{i,j}$ represents the $j_{th}$ element of feature vector of node $v_i$.

Now given a  network $G=(V,E)$, the problem is to predict whether a missing link edge $e' \in E'=  V \times V - E$ exists. 

\subsection{Circle Features}
To judge whether two nodes are connected with an edge,  we propose circled features to figure out whether two nodes are connected by circles of friends. The hypothesis is that if two nodes are connected by more circles with smaller size, the higher the probability that two nodes are connected is.

\subsubsection{Modified Swing Feature}

\begin{figure*}[htbp]
\centering
\includegraphics[width=100mm,scale=0.5]{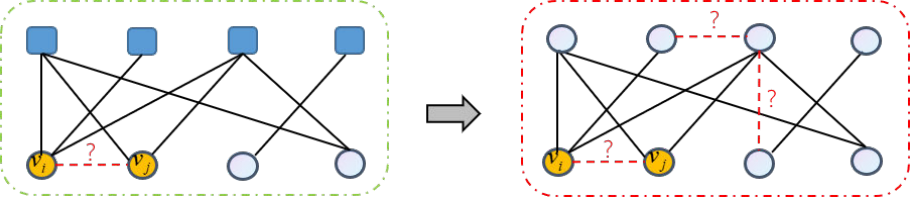}
\caption{The first circle feature: From Swing to Swing Plus.} \label{fig_swing}
\end{figure*}

The first proposed circled feature is a modified swing feature adapted to common graph, which is derived from an earlier work that can be applied only in bipartite graph \citep{yang2020large}. For common graph, missing links for any two nodes should be predicted, which is different from the original swing method. See Figure \Ref{fig_swing}. For original swing, any missing links between item nodes can be predicted. To differentiate link strength for any two nodes, we define a modified swing feature to rewrite the original adjacency matrix.

$$
c_{ij}^{swing+} = a_{i,j} + \sum_{u \in \Gamma_{i} \cap \Gamma_j}\sum_{v \in \Gamma_{i} \cap \Gamma_j}{\frac{1}{ \alpha + \mid \Gamma_u \cap \Gamma_v \mid }}
$$
, where $a_{ij}$ is an element of adjacency matrix, $\Gamma_{i}$,  $\Gamma_j$, $\Gamma_u$ and $\Gamma_v$ are common neighbors of corresponding nodes $i$,$j$,$u$ and $v$, and $\alpha$ is a constant.

\subsubsection{Bridge Feature}

\begin{figure*}[htbp]
\centering
\includegraphics[width=100mm,scale=0.5]{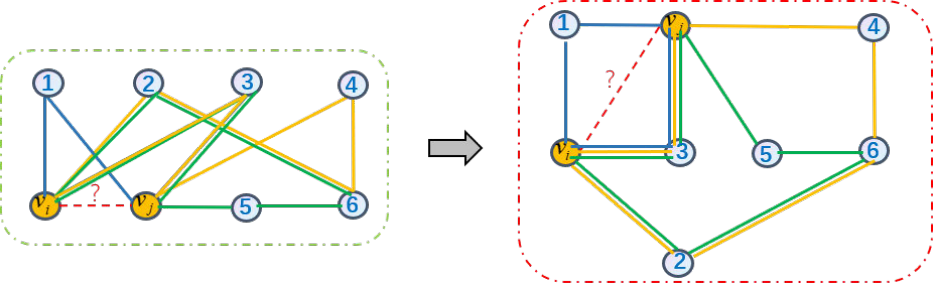}
\caption{Overview of the second circle feature: Bridge.} \label{fig_bridge}
\end{figure*}

The second proposed circle feature is bridge, which indicates the number of bridges that share the same circle. See Figure \Ref{fig_bridge}. Nodes $v_i$ and $v_j$ share 3 circles. Specifically, we give the computation formula for bridge as follows.

$$
c_{ij}^{bridge} = a_{i,j} + \frac{1}{2} \tanh(c) + \sigma(c) - 0.5
$$, where $\sigma$ is Sigmoid function, and $\tanh$ is hyperbolic tangent function, and $c$ is number of bridges sharing the same circle that cover nodes $i$ and $j$.

\subsection{Graph Self-attention Mechanism}
Based on SIEG\citep{sieg2023} network, which is constructed with a double tower structure of GCN\citep{kipf2016semi} and Graphormer \citep{ying2106transformers}, we propose circle features to stimulate graph self-attention mechanism such that graph transformer can be improved.
Specifically, to capture local graph structure information, we introduce the circle features to the transformer attention layer. Formally, we define the attention between two centered nodes as follows.
$$
A_{ij}^t = \frac{Q_i K_j^T}{\sqrt{d}} + b_{ij}^{dis} + b_{ij}^{num} + f_{ij}^{AA} + f_{ij}^{Jac} + {\color{red} c_{ij}^{swing+}} + {\color{red}c_{ij}^{bridge}}
$$
, where $Q_i$ is the $i_{th}$ row of Query matrix, $K_j$ is the $j_{th}$ row of Key matrix, $d$ is the dimension of $Q_i$ and $K_j$, $b_{ij}^{dis}$ is the distance of the shortest path between nodes $i$ and $j$, $b_{ij}^{num}$ is the number of the shortest path between the two nodes, $f_{ij}^{AA}$ and $f_{ij}^{Jac}$ are the common neighbor status with the Adamic-Adar form and the Jaccard form respectively. ${\color{red} c_{ij}^{swing+}}$ and ${\color{red}c_{ij}^{bridge}}$ are our proposed circle features.

\section{Experiments and Results}
\label{sec:exp}
Below we evaluate the performance of our method on OGB dataset ogbl-citation2\citep{hu2020ogb}.

\subsection{Datasets and task} As mentioned before, the ogbl-citation2 dataset is a directed graph, representing the citation network between a subset of papers extracted from MAG \citep{wang2020microsoft}. Each node is a paper with 128-dimensional word2vec features \citep{mikolov2013distributed} that summarizes its title and abstract, and each directed edge indicates that one paper cites another. Each node additionally has meta-information that identifies the year the associated paper was published. Given the citations already present, the aim is to predict missing citations. To be more specific, two references from each source paper are dropped at random, and a model is anticipated to place the two references that are missing higher than potential negative references.

\subsection{Baselines} We compare our method CFG with PLNLP~\cite{wang2021pairwise}, AGDN w/GraphSAINT~\cite{sun2020adaptive}, SEAL~\cite{zhang2018link}, S3GRL(PoS Plus)\citep{louis2023simplifying}, SUREL\citep{yin2022algorithm}, NGNN + SEAL \citep{DBLP:journals/corr/abs-2111-11638} and SIEG\citep{sieg2023}. 

\subsection{Metrics and settings} The evaluation metric is Mean Reciprocal Rank (MRR), where the reciprocal rank of the true reference among the negative candidates is calculated for each source paper, and then the average is taken over all source papers. The experiments are executed in PyG with Pytorch under the circumstance of Tesla A100 GPU(80G RAM).

\subsection{Experimental Results}
For comparison with existing methods, the performance of baseline methods are extracted from the ogbl-citation2 leaderboard and listed in the upper part of Table ~\ref{tab_citation2_result}. In the lower part of the table, the results of CFG are figured out by conducting our method CFG ten times in the same environment with the same associated settings. $CFG_1$ represents the method CFG using the first proposed circle feature swing plus, while $CFG_2$ represents the one using both the first and the second circle features, namely swing plus feature and bridge feature.
CFG beats all of the listed methods. It verifies the effectiveness of CFG model especially on dataset ogbl-citation2.

\begin{table}[htbp]
    \centering
    \caption{Performance of GNN models on dataset ogbl-citation2.}\label{tab_citation2_result}
    \begin{tabular}{lccc}
    \hline
   Method                                    & Test MRR  & Validation MRR             \\ \hline
    PLNLP    & 0.8492 ± 0.0029          & 0.8490 ± 0.0031          \\
    AGDN w/GraphSAINT  & 0.8549 ± 0.0029          & 0.8556 ± 0.0033         \\
    SEAL                                & 0.8767 ± 0.0032          & 0.8757 ± 0.0031         \\
    
    S3GRL (PoS Plus)   &0.8814 ± 0.0008 & 0.8809 ± 0.0074 \\
    SUREL & 0.8883 ± 0.0018         & 0.8891 ± 0.0021        \\
    NGNN + SEAL       & 0.8891 ± 0.0022         & 0.8879 ± 0.0022          \\
    SIEG   & 0.8987 ± 0.0018 & 0.8978 ± 0.0018 \\  \hline
    $CFG_1$   & \textbf{0.8997 ± 0.0015} & \textbf{0.8987 ± 0.0011} \\
    \hline
    \end{tabular}
\end{table}

\section{Conclusion}  \label{section_conclusion}
In this paper, we propose a Circle Feature aware Graph transformer(CFG) for link prediction task, which can stimulate graph transformer. CFG achieves top 1 performance on dataset ogbl-citation2 until submission. The experimental results verify the superiority of CFG. We will optimize transformer self-attention mechanisms with circle features in future work.

\bibliographystyle{IEEEtran}
\bibliography{mybibliography}
\end{document}